\documentclass[review]{elsarticle}
\usepackage{amssymb}

\journal{Chaos, Solitons and Fractals}

\begin{document}

\begin{frontmatter}
\title{Efficient synchronization of structurally adaptive coupled Hindmarsh-Rose neurons}
\author{A. Moujahid, A. d'Anjou, F. J. Torrealdea}
\address{Department of Computer Science, University of the
Basque Country, UPV/EHU 20018 San Sebastian, Spain}

\author{F. Torrealdea}
\address{UCL Institute of Neurology, Queen Square-London WC1N 3BG}

\date{\today}

\begin{abstract}
 The use of spikes to carry information between brain areas implies complete or partial synchronization of the neurons involved. The degree of synchronization reached by two coupled systems and the energy cost of maintaining their synchronized behaviour is highly dependent on the nature of the systems. For non-identical systems the maintenance of a synchronized regime is energetically a costly process. In this work, we study conditions under which two non-identical electrically coupled neurons can reach an efficient regime of synchronization at low energy cost. We show that the energy consumption required to keep the synchronized regime can be spontaneously reduced if the receiving neuron has adaptive mechanisms able to bring its biological parameters closer in value to the corresponding ones in the sending neuron.
\end{abstract}

\begin{keyword}
Adaptive synchronization, Synchronization energy 
\end{keyword}

\end{frontmatter}

\maketitle

\section{Introduction}
\label{sec:1}

Neural information processing depends on communication between neurons that involves partial or complete synchronization in their signalling. When a neuron receives signals from its neighbours its firing regime undergoes changes that lead to a modification of its information capacity as well as its average energy consumption. The degree of synchronization reached by two neurons conditions their capacity to exchange information and the energy cost to keep their synchronous signalling activity. Two structurally similar coupled systems can reach different degrees of synchronization depending on whether   they are identical or not.  Identical systems reach complete synchronization spontaneously beyond a given value of the coupling strength and the energy required to maintain this completely synchronized regime is zero. However, between non-identical systems complete synchronization never occurs spontaneously without cost and an average nonzero flow of energy is required to maintain a completely synchronized regime \cite{1}.
In this work we investigate, in terms of energy consumption, how an efficient degree of synchronization between two non-identical electrically coupled Hindmarsh-Rose neurons can be reached. We show that feedback synchronization at sufficiently large coupling force between two non-identical neurons creates appropriate conditions for efficient actuation of adaptive laws able to make the neurons approach each other their biological parameters values in order to decrease the energy consumption.
In the next section, we describe the four dimensional Hindmarsh-Rose neuron model used to represent the dynamics of real neurons, and we report an energy function associated to that model. The analysis of energy consumption during the synchronization process between the two neurons, and the description of the adaptive laws are performed in Section 3. Numerical results are showed and analyzed in section 4. Finally, in Section 5 we collect the main conclusions of this work.

\section{The Hindmarsh-Rose neuron energy}
\label{sec:2}

In this paper, we represent a single neuron by the four-dimensional Hindmarsh-Rose neuron model described by the following equations of movement \cite{2,3,4,5}:
\begin{equation}
\begin{array}{l}
            \dot{x}=ay+bx^2-cx^3-dz+\xi I,    \\
            \dot{y}=e-fx^2-y-gw,      \\
            \dot{z}=m(-z+s(x+h)), \\
 	    \dot{w}=n(-kw+r(y+l)),
\end{array}
\label{equ1}
\end{equation}
where $a,b,c,d,\xi,I,e,f,g,m,s,h,n,k,r$, and $l$ are the parameters that govern the dynamics of the neural system. The variable $x$ is a voltage associated to the membrane potential, variable $y$ although in principle associated to a recovery current of fast ions has dimensions of voltage, variable $z$ is a slow adaptation current associated to slow ions, and variable $w$ represents an even slower process than variable $z$ \cite{6}. $I$ is a external current input.

In a previous work, we have assigned an energy function to the Hindmarsh-Rose model given by Eq. (\ref{equ1}). The procedure followed to find this energy has been reported in detail in \cite{7}.  This energy function $H(\mathbf x)$ is given by
\begin{equation}
\begin{array}{ll}
 H=\frac{p}{a}(\frac{2}{3}fx^3+\frac{msd-gnr}{a}x^2+ay^2) \\
+\frac{p}{a}(\frac{d}{ams}(msd-gnr)z^2-2dyz+2gxw)
\label{equ2}
\end{array}
\end{equation}
In the model time is dimensionless and
every adding term in Eq. (\ref{equ2}) has dimensions of square voltage, so
function H is dimensionally consistent with a physical energy as
long as parameter $p$ has dimensions of conductance. In this paper we fix parameter $p$ to the arbitrary value $p=-1 S$. We have adopted a negative sign for this parameter so that the outcome of the model will be consistent with the usual assumption of a demand of energy associated with the repolarization period of the membrane potential and also with its refractory period.

The energy derivative $\dot{H}$ is given by
\begin{equation}
\dot{H}= \frac{2p}{a}\left( \begin{array}{c}
            fx^2+\frac{msd-gnr}{a}x+gw     \\
            ay-dz      \\
            \frac{d}{ams}(msd-gnr)z-dy \\
 	    gx
           \end{array}
    \right)^T
\left( \begin{array}{c}
            bx^2-cx^3+\xi I     \\
            e-y      \\
            msh-mz \\
 	    nrl-nkw
           \end{array}
    \right)
\label{equ3}
\end{equation}
which is also dimensionally consistent with a dissipation of energy.

 The energy derivative given by Eq. (\ref{equ3}) represents the net energy variation over time of the energy of the neuron. For an isolated neuron, whose dynamics is confined to an attractive region of the state space, its average net variation of energy is zero what indicates that the energy the neuron obtains through the membrane is perfectly balanced by its dissipation of energy. Nevertheless, when the neuron is forced to synchronize to another neuron its balance of energy through the membrane is broken. In this case, the coupling device must counterbalance the net energy flow through the membrane in order to lead to a zero global average variation. For the particular system given by Eq.  (\ref{equ5}), the net variation of energy at the coupling device is given by the following expression according to Ref. \cite{8}, where the first vector stands for the gradient of the energy function given by Eq. (\ref{equ2}) and the second term represents the coupling device (see Eq. (\ref{equ5})). $T$ denotes transpose of a matrix.

\begin{equation}
\frac{2p}{a} \left( \begin{array}{c}
            fx^2+\frac{msd-gnr}{a}x+gw     \\
            ay-dz      \\
            \frac{d}{ams}(msd-gnr)z-dy \\
 	    gx
           \end{array}
    \right)^T
\left( \begin{array}{c}
            \mathbf {k}(x_1-x_2)     \\
            0      \\
            0 \\
 	        0
           \end{array}
    \right)
\label{equ4}
\end{equation}

\begin{figure}[ht]
\begin{center}
\includegraphics[width=1\textwidth]{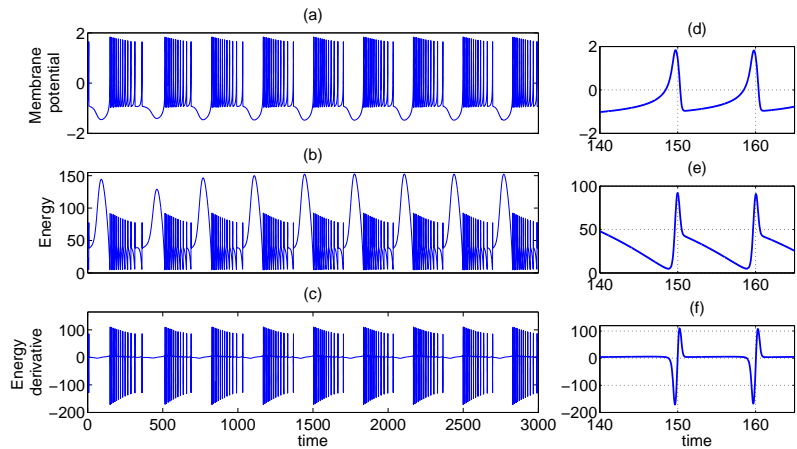}
\caption{(a) Action potentials, (b) energy and (c) energy derivative for the Hindmarsh-Rose model neuron. (d), (e) and (f) Details of the action potential, energy and energy derivative associated to two spikes}
\end{center}
\label{fig1}
\end{figure}

For the isolated neuron of Eq. (\ref{equ1}), we have computed both energy and energy derivative corresponding to a series of action potentials. The results are depicted in Figure 1.
Fig. 1 (a) shows a series of action potentials (variable $x$ in the model neuron). Fig. 1 (b) and Fig. 1 (c) show both energy and energy derivative corresponding to that action potentials. Fig. 1 (d) shows two action potentials. Fig. 1 (e) and Fig. \ref{fig1} (f) show  detail of energy and energy derivative associated to a train of two action potentials. The energy values are negative but have been depicted  in Fig. 1 (and later in figures 2 and 5) with a positive sign for a better appreciation of the energy demand when generating a spike.
For each action potential it can be appreciated that the energy derivative is first negative, dissipation of energy while the membrane potential depolarizes during the rising phase of the spike, and
then positive, contribution of energy to repolarize the membrane potential during its falling phase. During the refractory period
between the two spikes the energy derivative remains slightly positive, still demanding energy, until the onset of the following action
potential.

In the next section, we analyze the average of energy and energy derivative of two nonidentical coupled Hindmarsh-Rose neurons. We consider that, in a first stage, the coupled neurons have been forced to synchronize with each other, then we initiate an adaptive process that adapts the mismatched parameters of the receiving neuron to the one of the sending neuron.

\section{Adaptive mechanism}
\label{sec:3}

Gap junction channels permit the intracellular electrical potentials of two neurons to directly connect together and are usually referred to as electrical synapses. They are frequent when two or more neurons are coupled together and play an important role in the synchronization of cellular events. In particular, they are efficient in transmitting information and in synchronizing the information of groups of neurons \cite{9}. Although symmetry is an expected property for an electrical synapse, asymmetric gap junctions have also been reported in
the literature \cite{10}. In this section we analyze the energy requirements involved in achieving a complete synchronization between two initially nonidentical Hindmarsh-Rose neurons. We first consider an asymmetric unidirectional coupling scheme in which only the first variable of the receiving neuron is affected by the coupling according to the following system of equations:

\begin{equation}
\begin{array}{l}
            \dot{x}_i=ay_i+bx_i^2-c_ix_i^3-dz_i+\xi I_i+\mathbf k_i(x_j-x_i), \\
            \dot{y}_i=e_i-f_ix_i^2-y_i-gw_i,      \\
            \dot{z}_i=m(-z_i+s(x_i+h)), \\
 	    \dot{w}_i=n(-kw_i+r(y_i+l)),
\end{array}
\label{equ5}
\end{equation}
where $\mathbf k_1 = 0$ and $\mathbf k_2 \geq 0$ is the coupling strength. Note that the coupling affects only the membrane voltages $x_2$ of the receiving neuron. $i,j=1,2;i\neq j$ are the indexes for the neurons.

We suppose that the sending neuron is set in the chaotic spiking-bursting regime corresponding to a constant external current $I_1=3.024$, with the standard parameter values $a = 1$, $b = 3.0 (mV)^{-1}$, $c_1 = 1 (mV)^{-2}$, $d = 0.99M\Omega$, $\xi = 1M\Omega$,
$e_1 = 1.01mV$, $f_1 = 5.0128 (mV)^{-1}$, $g = 0.0278M\Omega$, $m = 0.00215$, $s = 3.966\mu S$, $h = 1.605 mV$, $n = 0.0009$, $k = 0.9573$, $r = 3.0\mu S$, $l = 1.619mV$.\\

The receiving neuron is initially set to its quiescent state at a low value $I_2=0.85$ of its external current, the parameters $c_2$, $e_2$ and $f_2$ are mismatched as follow: $c_2= 0.95 (mV)^{-2}$, $e_2 = 0.85mV$, $f_2 = 5.1128 (mV)^{-1}$. All others parameters are the same for both neurons.

Under these conditions the coupled neurons are not identical and therefore synchronization does not spontaneously occur at any value of the coupling strength but, rather, it must be strongly enforced through the establishment of large values of the coupling strength which will involve high energy consumption. The energy consumption required to maintain the synchronized regime can be spontaneously reduced if the receiving neuron is flexible enough as to adapt its parameters through an adequate adaptive law in order to reach the nominal value of the sending neuron parameters.
 Ideally, if the neurons become identical their joint dynamics is attracted toward a regime of zero error in the variables. This asymptotical limit regime of identical synchronization occurs with zero net average exchange of energy with the environment.

In a previous work \cite{8} we deduced adaptive laws that permit to any family of homochaotic coupled systems to adapt their structures in order to approach each other provided they are previously forced to a certain degree of synchronization.

The coupling scheme given by Eq. (\ref{equ5}) can be written, in general, as follow:

 \begin{equation}
\begin{array}{l}
            \dot{\mathbf x_1}=f(\mathbf x_1,\mathbf p),\\
            \dot{\mathbf x_2}=f(\mathbf x_2,\mathbf q)+\mathbf K(\mathbf x_1-\mathbf x_2),      \\
\end{array}
\label{equ6}
\end{equation}
where $\mathbf x_1,\mathbf x_2\in \Re^n$ indicate the states of the coupled neurons,  $\mathbf p$ and $\mathbf q$ stand for the parameters of the sending and receiving neurons,  and  that $\mathbf K \in \Re^n$ is a diagonal matrix representing the coupling strength with entries $\mathbf k_i \geq 0, i=1,...,n$. In the particular case where only the first variable of the vector  $\mathbf x_2$ is affected by the coupling, we have $\mathbf k_1=\mathbf k$ and $\mathbf k_i=0, i=2,...,n$.

If the coupling strength $\mathbf k$ is large enough as to make the errors in the variables $\zeta=\mathbf x_2-\mathbf x_1$ small, an operational law that adapts the parameters of the receiving neuron to the ones of the sending neuron is given by \cite{8}

 \begin{equation}
\begin{array}{ll}
            \dot{\zeta}_i^p=- \left[ \sum_{l=1}^n \left( \frac{\partial f_l(\mathbf x_2,\mathbf q)}{\partial \mathbf q_i} \right) _{(\mathbf x_1,\mathbf p)} \zeta_l \right]\\
            - \left[\sum_{l=1}^n \sum_{j=1}^n \left( \frac{\partial^2 f_l(\mathbf x_2,\mathbf q)}{\partial \mathbf q_i \partial \mathbf x_2^j} \right)_{(\mathbf x_1,\mathbf p)} \zeta_j \zeta_l    \right]
\end{array}
\label{equ7}
\end{equation}
where $\zeta^p=\mathbf q-\mathbf p$ denotes the vector of parameter errors, and the summation is over every component of the vector field $f$. The above law is general and can be used to find specific adaptive laws to any kind of homochaotic systems provided they are coupled through a feedback scheme of large enough coupling strength.

Based on the above adaptive mechanism, and according to  Eq. (\ref{equ7}), the adaptive laws that govern the dynamics of the mismatched parameters of the receiving neuron of the system of Eq. (\ref{equ5}), are given by the following equations:

 \begin{equation}
\begin{array}{l}
            \dot{I}_2=- \xi (x_2-x_1)\\
            \dot{c}_2=x_1^3 (x_2-x_1)+3x_1^2(x_2-x_1)^2 \\
            \dot{e}_2=-(y_2-y_1) \\
            \dot{f}_2=x_1^2 (y_2-y_1)+2x_1(x_2-x_1)(y_2-y_1) \\
\end{array}
\label{equ8}
\end{equation}

\section{Numerical results}
\label{sec:4}
Numerical results have been performed simulating the system of Eq. (\ref{equ5}) over $50 000$ time units.

In a first phase,  we force the receiving neuron to replicate the behavior of the sending neuron through a progressive increase of the coupling force $\mathbf k$ to a value large enough to ensure a regime close to complete synchronization. When $\mathbf k$ reaches the appropriate value we keep it fixed for a second phase in which we begin the process of parameters adaptation. During the forced synchronization the parameter $\mathbf k$ has been increased linearly from $0$ to $30$, value that has remained fixed during the adaptation phase. The registrated values corresponding to energy and energy derivative have been averaged over a convenient length of time in order to avoid large fluctuations. The dissipated energy has been averaged over five units of time, while the proper energy has been averaged over ten units of time. The energy at the coupling device is calculated according to Eq. (\ref{equ4}).

\begin{figure}[ht]
\begin{center}
\includegraphics[width=1\textwidth]{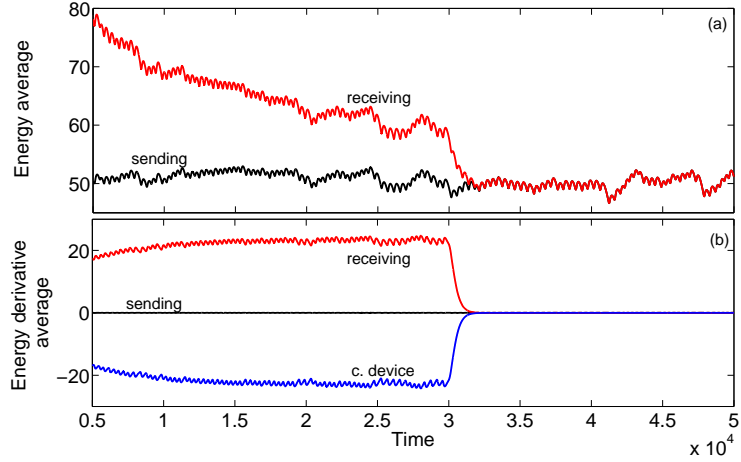} \end{center}
\caption{In (a) Average energy over five units of time of sending and receiving neurons. In (b) average per unit time of the energy variation over ten units of time.  Adaptation begins at $t=30,000$.}
\label{fig2}
\end{figure}

Figures \ref{fig2}(a) reports the average values of energy per unit time of sending and receiving neurons. We can see that in a first stage ($t<30,000$), the receiving neuron is forced to synchronize with the sending neuron, and shows a decreasing pattern of average energy  due to the gradual increase of the coupling force. 
In this regime, the receiving neuron  is continuously demanding energy from the coupling device, leading to an increase of the energy variation through the membrane as we can see in part (b) of the Fig.\ref{fig2}. The energy demanded by the receiving neuron must be supplied by the coupling device.
After adaptation takes place, the two neurons become structurally close each other, and enter in a completely synchronized regime of zero error in the variables as shown in Fig. \ref{fig4}. This regime corresponds to a balanced exchange of energy between the neuron and its environment corresponding to a zero value of the energy derivative.

The energy consumption of the sending neuron remains constant because its dynamics is not affected by the coupling, this energy is perfectly balanced by the income of
energy it receives through the membrane, i.e. its energy variation is  $\dot{H}_1=0$.

\begin{figure}[ht]
\begin{center}
\includegraphics[width=1\textwidth]{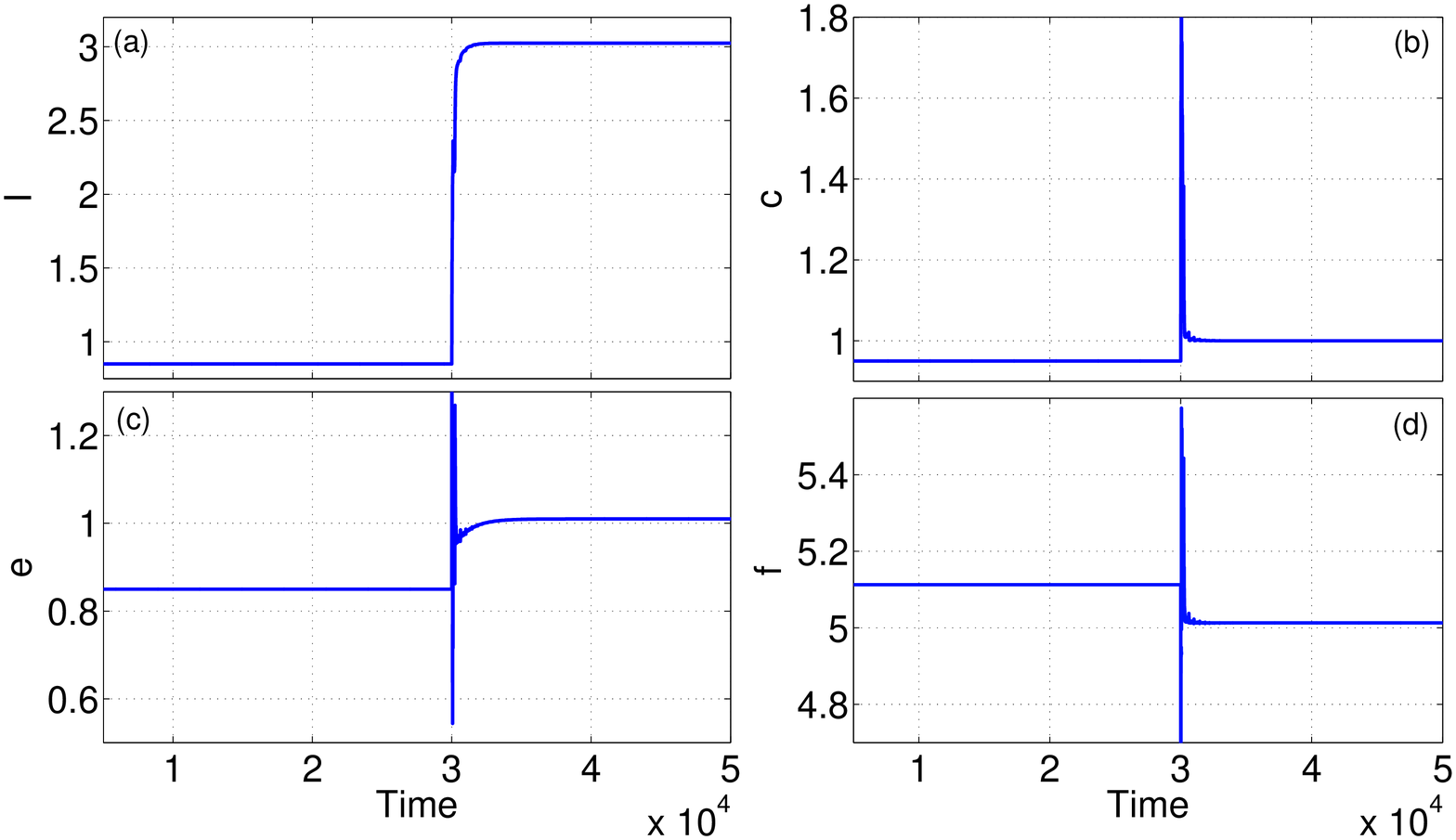} \end{center}
\caption{In (a)-(d) adaptation results of the mismatched parameters of the receiving neuron.}
\label{fig3}
\end{figure}

\begin{figure}[ht]
\begin{center}
\includegraphics[width=1\textwidth]{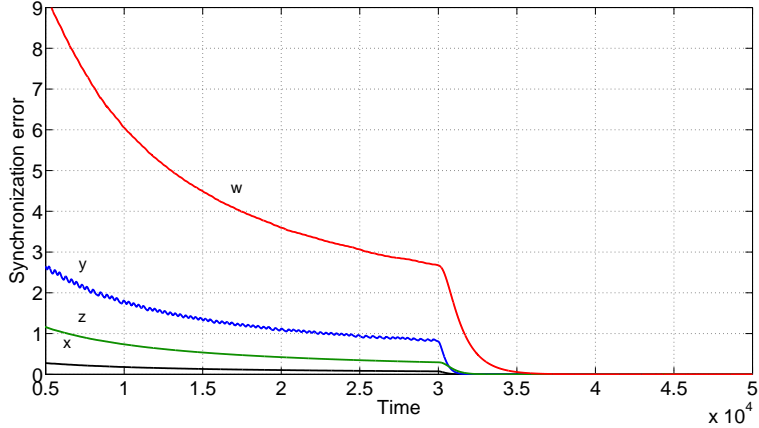} \end{center}
\caption{Synchronization errors in state space variables.}
\label{fig4}
\end{figure}

In Fig. \ref{fig3} we report the results of the adaptation mechanism that has been implemented according to Eq. (\ref{equ8}) to adjust the mismatched parameters of the receiving
neuron to those of the sending neuron. As it can be appreciated, a correct adaptation of the accessible parameters is achieved. In Fig. \ref{fig4}, we can see that during the first phase, before the adaptation process begins, the two neurons have already experienced a notably decrease in the synchronization error which correspond to a high level of synchronization. This synchronized signalling implies that information is efficient in the sense that it is being transmitted between both neurons at high signal to noise ratio. However, as Fig.  \ref{fig2}(b) shows, to keep this synchronised regime implies a high energy cost to the coupling device. In other words, maintaining the synchronized regime for communication has an additional energy cost over the basic energy cost of the normal neuron's metabolism. This additional cost is direct consequence of the two neurons being non-identical. Figure \ref{fig2}(b) shows in a second phase that, if the receiving neuron adapts its biological parameter values, the two neurons synchronize with zero net flow of energy from the coupling device. That is, the efficiency of the information transmission is slightly increased and transmission itself has no energy cost.

As bidirectional couplings between neurons are frequent, we have also considered a bidirectional electrical coupling corresponding to $\mathbf {k}_1=\mathbf {k}_2=\mathbf {k}$ in Eq. (\ref{equ5}). In this case, the mutual interaction induced by the coupling affects both the variable $x_1$ and $x_2$ of the sending and receiving neurons. As in the previous experiment, we analyze the energy consumption required to maintain the synchronized regime between the coupled neurons, and the energy average that they receive from the external source, i.e., the coupling device.

We suppose that the dynamics of the sending neuron is only affected by the
coupling but not by the mechanism of parameters adaptation, so that all its parameter values are kept fixed whereas the mismatched parameters of the receiving
neuron are governed by the adaptive laws given by Eq. (\ref{equ8}).\\

\begin{figure}[ht]
\begin{center}
\includegraphics [width=1\textwidth]{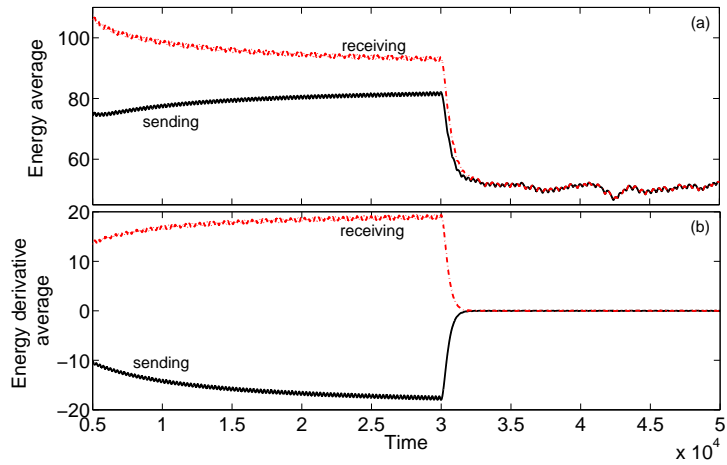} \end{center}
\caption{In (a) energy average over five units of time of the sending and the receiving neurons. In (b) average per unit time of the energy variation over ten units of time of the sending and the receiving neurons.  Adaptation begins at $t=30,000$.}
\label{fig5}
\end{figure}

Figure \ref{fig5} shows the average values of energy and energy derivative per unit time of the sending and receiving neurons.

As it can appreciated, before adaptation starts ($t < 30, 000$), and as soon as the coupling device is connected, the coupled neurons begin to adjust their dynamics to achieve a common behaviour. At this stage, the sending neuron, as shown in Fig. \ref{fig5}(b), undergoes an increase in its average energy which corresponds to an increase of its dissipation of energy (negative values of its energy derivative). On the other hand, the receiving
neuron displays an opposite pattern corresponding to a decrease in its average energy (see Fig. \ref{fig5}) and an increase in its demand of energy to be supplied by the coupling device. After adaptation takes place, the neurons reduce notably their average energy values to reach a value of about 50 corresponding to the average energy of an isolated neuron (see Fig. \ref{fig5}(a)). At this stage the energy variation of both neurons is zero.

\section{Conclusion}

Signalling in synchrony is a normal way to propagate information between neurons. However, for non-identical neurons the cost of signalling in synchrony is superior to the mere metabolic energy cost that the neurons would have if they were signalling independently.  Maintaining a synchronized regime between two non-identical neurons requires energy that must be provided by the coupling biological device that forces that synchrony. In contrast with non-identical neurons, identical neurons could reach a spontaneous synchronized regime at zero maintenance energy cost.\\

This paper analyzes the energy cost of maintaining a synchronized regime between two electrically coupled Hindmarsh-Rose neurons. We have considered a sending neuron always signalling in a chaotic regime and a non-identical receiving neuron, with some mismatch in its parameters, initially set in its quiescent state. Under these conditions, to keep the synchronized regime requires a net flow of energy that can be costly to maintain. We show how this energy cost reduces notably when the receiving neuron is able to adapt its structure in order to approach the dynamics of the sending neuron. Biological structures are particularly flexible in adapting their parameters and the mechanism of adaptation introduced in this work could make some of the required collective behaviours between groups of neurons energetically less costly.

\end{document}